# Information Technology Needs in Vehicle Dynamics Control

Levent Güvenç and Bilin Aksun Güvenç
*Department of Mechanical Engineering, İstanbul Technical University*
*E-mail: guvenc@mkn.itu.edu.tr*

**Abstract**

*Significant changes are occuring in the field of vehicle dynamics control. Consequently, vehicle dynamics control systems are expected to be as common as ABS systems in the near future. This paper focuses on the information technology related requirements posed by these advances in the area of vehicle dynamics control. Topics considered include vehicle dynamics simulation tools, hardware in the loop simulation systems, sensors and related fault tolerance/diagnostics, man machine interface problems and the implications of a networked architecture on controller design.*

## 1. Introduction

In recent years, the automotive sector has been making increasingly more use of electronic control systems. Anti-lock brakes (ABS) will soon become mandatory in Europe. Traction control systems are also present in most cars. Vehicle dynamics control (VDC) systems that utilize the ABS system for individual wheel braking and the traction control system for adjusting motor power are starting to become more readily available in commercial vehicles. Currently, every major automotive company and related supplier is working on improving their VDC algorithms that aim at making the average driver at least as skilled as a professional driver in stabilizing his car. VDC systems counteract undesired yaw motions of the vehicle during understeer or oversteer situations and unsymmetrical dynamic loading like braking on partially icy road, a side wind or a tire blowout. For instance, in μ-split braking where the vehicle enters an ice patch only on one side, the driver's panic reaction response is not fast enough to prevent the undesired and dangerous yaw rotation that results. Most drivers overreact in such situations, worsening the stability problem. In such situations, a well designed VDC will prevent the vehicle from losing stability. Work on incorporating roll-over avoidance schemes into VDC is currently in progress. Steer-by-wire systems, whereby the mechanical steering linkage will be obsolete, are expected to appear in the commercial market by 2002 and brake-by-wire systems will appear as early as 2004. Indeed, today's conventional cars are expected to become by-wire systems using mainly electromechanical actuators and a large array of sensors by the year 2010 (Bretz [14]). Some car manufacturers have already built concept cars to that effect. These exciting expected developments will force automotive manufacturers to use the synergistic approach of mechatronics. It will hence be easier to implement VDC algorithms but the complexity of the algorithms and the required fault tolerance and diagnostic error checking will impose increasingly more demanding requirements on the electronic control units (ECU) being used. VDC designers will have to spend a progressively increasing amount of time on meeting and sometimes outperforming information technology induced constraints.

This paper focuses on information technology needs in vehicle dynamics control. The paper gives an overview of information technology needs within certain sections of VDC. These sections are classified here as: vehicle dynamics simulation software and hardware; hardware-in-the-loop simulation; sensors, signal conditioning, fault tolerance and diagnostics; man-machine interface problems; and requirements on ECUs, networking requirements and their implications for controller design.

The paper starts with a survey of several different approaches to vehicle dynamics control. After reviewing each section and presenting the state-of-the-art, the paper concentrates on man-machine interface problems and the implications of a networked VDC structure on control design and performance in the last two sections. The man-machine interface problem is quite important as the action of the driver when coupled with the VDC can lead to problems of stability if this effect is not properly studied and used in the VDC algorithms. The approach used here does not allow the VDC to take over control of the car completely. Instead, the VDC takes over control only during the panic reaction time of the driver and lets control back to the driver after its initial compensatory action. Another important consideration in VDC algorithm development is that the ECU (which reads the feedback signals from the sensors, does all the required error checking, computes the stabilization algorithm and sends the necessary correction signals, if necessary) works over a network. Network delays should, thus, be paid attention to in the VDC algorithm design. Simulation studies are used in the last section to study the effect of such network delays. The paper ends with the usual conclusions and recommendations.

## 2. Vehicle dynamics control

Most of the existing work on vehicle dynamics controller development can be broadly divided into the two categories of automated path following and vehicle yaw stabilization. Extensive research work has been conducted in the US on automated path following due to the government backed desire to build Intelligent Vehicle Highway Systems (IVHS) whereby vehicles forming a platoon will travel automatically without requiring driver intervention (Shladover [50], Shladover et al [49], Byrne et al [43], Peng and Tomizuka [20], Guldner et al [32], Tan et al [23], Unyelioglu et al [34]). Successful operation of such vehicle platoons on test tracks have been reported. While being a very complicated problem when viewed with the mechatronics approach or when considering the information technology needs, at least, controller design is simplified (Tomizuka and Hedrick [38]) due to the absence of the driver in the longitudinal and lateral control loops. Automated path following has also attracted attention outside the US. Parts of the PROgramme for a European Traffic with a Highest Efficiency and Unprecedented Safety (PROMETHEUS) of the European Union deal with automated path following. Outside the US and Europe, Intelligent Transport Systems (ITS) have also found interest in Japan (see Nwagboso [11] for more details of IVHS and ITS). An interesting application of automated path following involves automated lane following of buses on separate narrow lanes (Darenberg [56], Ackermann and Darenberg [27], Ackermann et al [30], Aksun Güvenç and Güvenç [5]). The possibility of using narrow lanes for buses reduces costs and such a system has been in operation in Essen, Germany since 1995.

The second category of vehicle yaw stabilization has found more interest outside the US, mainly in Europe and in Japan. Vehicle yaw stabilization controllers which are also called Vehicle Dynamics Controller (VDC) or Electronic Stability Program (ESP) make sure that the vehicle does not exhibit undesired yaw rotations (spin motions) due to undesirable tyre-road interaction. They intervene during unsymmetrical loading like in μ-split braking or side wind forces. They are also used to avoid understeer or oversteer when taking turns. One out of every three cars in Europe is expected to have a VDC system by 2004. In vehicle yaw stabilization, the driver is in the feedback loop which means that a driver model is needed for controller design, which, of course, complicates the design procedure. The controller used has to have additional robustness to accommodate for different driver characteristics. One very simple solution that is used in Section 6 here is to allow short duration control intervention, only when necessary, and to let control back to the driver at other times.

Automated path following controllers are necessarily steering controllers for vehicle lateral control. Throttle and brake control are also used in automated path following for longitudinal control of the vehicle in the platoon. On the other hand, yaw stabilization controllers are not necessarily steering controllers. Any practical way of producing the necessary corrective yaw moments is acceptable as an actuation system. Correspondingly, there are two approaches to hardware implementation of vehicle yaw stabilization controllers. The more commonly used one is to achieve yaw motion control through individual braking of the wheels. The presence of the ABS actuators simplifies the individual wheel braking implementation of vehicle dynamics control. This is the basis of the ESP system of Bosch (van Zanten [3]). The second approach is to achieve the required corrective yaw motions by using a steering controller. The advantages of using a steering controller are that steering is a continuous operation as compared to braking and that relatively shorter braking distance can be achieved (Ackermann [28]). Robust decoupling of steering and yaw stabilization by yaw rate feedback has been done by Ackermann [29]. Currently, one needs to add an additional controllable steering actuator, obtaining an auxiliary steering system with control saturation limitation, to the mechanical steering linkage to implement a steering controller. However, with the expected utilization of steer-by-wire systems whereby the mechanical steering linkage will be replaced by a steering actuator with no mechanical connection to the steering wheel starting in 2002 (Bretz [14]), the implementation of steering controllers will be as easy as the implementation of individual wheel braking controllers. Future VDC systems are, thus, expected to use a combination of steering and braking intervention for yaw dynamics stabilization. Note that while the main aim in automated path following is to follow the lanes in a platoon, such systems also need to use a yaw stabilization type controller for each vehicle as a subsystem.

Some interesting references in the area of vehicle dynamics control will be cited next. Mammar [47] has applied two degree-of-freedom $H_\infty$ optimization to vehicle lateral control. The model regulator has been used by Aksun Güvenç et al [8-9] and Bünte et al [52] for the same problem. Chen and Tomizuka [10] have designed nonlinear control algorithms for lateral control of commercial heavy vehicles. Qu and Liu [42] have concentrated on the effect of the nonlinear tire characteristics on vehicle lateral dynamics. Fukada [54] and Hac and Simpson [1] have worked on estimating slip angle for vehicle lateral stability control.

The area of vehicle dynamics control is an ever growing one, with new capabilities being continuously investigated and incorporated into it. For example, VDC systems for light and heavy commercial vehicles are also being developed. When considering heavy commercial vehicles, one is concerned more with avoiding vehicle roll-over as compared to yaw stabilization alone (see Ma and Peng [53] for example). Roll-over avoidance is an important consideration for vehicles with high centers of gravity (quite necessary even for automobiles in countries with rough roads). Accordingly, there is a vast amount of research publications in the area of roll-over avoidance (Palkovics et al [35], Eger and Kiencke [44], Larson et al [46], Ackermann and Odenthal [24-25]). Note that the steering, braking and throttle systems had been mentioned until this point. In regard to roll-over, the suspension system of the vehicle is also very important. Semi active and active suspension systems have therefore also been the focus of investigations (see Hrovat [12] and the references therein). Properly designed active suspension controllers can be incorporated into VDC systems to minimize undesired vehicle roll while turning a corner. Another useful feature of semi active and active suspension systems is that ride and handling properties of the vehicle can both be optimized rather than finding a compromise between these two conflicting objectives.

## 3. Vehicle dynamics simulation software and hardware

The adequate development of the vehicle dynamics control algorithms in the previous section rely upon vehicle models of adequate complexity for controller design and highly realistic models for virtual testing of the designed controllers. While control engineers have prepared their simplified car simulation model in programs like Matlab/Simulink, multi-body dynamics program developers have also prepared special code for automotive applications. An early review of multibody computer codes for vehicle system dynamics carried out by Kortüm [57] gives a list of a large number of programs at that time. While some of those have vanished from the market, some new ones have also appeared. One of the most widely used ones in vehicle simulation is Adams (see Orlandea [40] for an earlier description). While the basic idea of Adams has not changed significantly since then, the increased computational power available now has made fast simulation of full vehicle motion, interactive 3-D graphics visualization and co-simulation with other programs like Matlab/Simulink possible. Adams now has a dedicated toolbox called Adams/Car for automotive applications. Adams/Car is a high end product and along with its use in vehicle design it also allows vehicle dynamics control engineers to apply their controller to a virtual vehicle before more expensive hardware-in-the-loop or road tests. Since programs like Adams/Car are also used in vehicle design (like the design or optimization of a virtual prototype vehicle's suspension) their use allows the control engineers to start testing their controllers during product development, enabling them to interact with the design engineers from the beginning rather than waiting until the final manufactured prototype. The increased synergy of the mechatronics approach is thus utilized by doing vehicle design and controller design simultaneously.

Control engineers prefer to analyze, design and build their controllers in specialized programs like Matlab/Simulink while their virtual plant exists in the separate vehicle dynamics simulation program. Co-simulation of both programs are necessary and almost all available programs allow such co-simulation (Villec [17]). In a typical co-simulation of Matlab/Simulink and Adams/Car, for example, all controller functions exist within Simulink and the plant under control is the car model in Adams/Car. In vehicle dynamics control, certain car variables like yaw rate, longitudinal speed and steering wheel angle have to be provided to the controller in Simulink which calculates the necessary control action and sends the steering actuator or individual wheel brake command to Adams/Car. Currently, this kind of co-simulation works (see Türker [55]) but is not problem free. The presence of dry friction, for example, can cause serious problems of accuracy and convergence. The information technology needs in this aspect of vehicle dynamics control are easy-to-use and easy-to-build co-simulation environments between control system simulation and vehicle dynamics simulation programs. In addition, these co-simulation environments have to be optimized for speed. It is also very important to have access to benchmark virtual cars (in programs like Adams/Car) such that independent researchers can test and validate their new vehicle dynamics control designs without always having to find an interested partner in the automotive sector.

With the ever increasing need to simulate more complicated maneuvers at less time, speed improvements will always be the number one information technology related need in vehicle dynamics simulation. Such speed improvements depend not only on better written and better optimized software but to a great extent, also on computational hardware. It is expected that the on-going race for faster, more powerful computers will have a positive influence on this need.

## 4. Hardware in the loop simulation

The ultimate test of a vehicle dynamics controller is done on a proving ground where a series of road-tire loads (wet road, icy road etc) and handling tests are conducted. Even after conceptual development of a vehicle dynamics control system using co-simulation as explained in the previous section, one may want to reduce prototype manufacture and test costs by confining hardware tests to a specific subsystem under development like the active steering or active suspension system. A hardware-in-the-loop (HIL) simulation testing is then commonly used in vehicle dynamics controller development where only the control actuation system exists in hardware in an otherwise soft simulation environment. The first information technology specific requirement for HIL equipment is high speed of operation as the software part of the loop has to be simulated at real time speed since the hardware part necessarily operates at real time. The second information technology specific requirement is the need to interface with automotive hardware. Based on these needs, the HIL software runs on a separate PC card with its fast processor. The vehicle dynamics model running on it has to be optimized for speed to be able to run at real time. The PC cards that are used are sometimes replaced by dedicated computers and sometimes several dedicated computers are used simultaneously during HIL testing (Schütte et al [21], Sauer and Gromeier [45], Zerfowski [13]). Due to the need to interface with automotive hardware running on a network, network protocol software has to be included along with enough digital and analog inputs and outputs for interfacing.

In HIL for vehicle dynamics controller development, the electronic control unit (ECU) being developed can easily be tested through an extensive range of maneuvers in the HIL environment (Pfister et al [41]).

## 5. Sensors, signal conditioning, fault tolerance and diagnostics

Sensors play a very important role in vehicle dynamics control. The most important sensor in VDC is the yaw rate sensor. Advances in manufacturing technology have resulted in micromechanical yaw rate sensors with all mechanical parts and electronics packed in one IC chip (Mörbe and Illing [37], Ichinose and Terada [54], Shkel et al [2]). These sensors use a micromachined gyroscope to measure yaw rate. Similarly, roll rate sensors (Mörbe and Illing [37]), steering wheel angle sensors (Gruber [31]) and accelerometers (Masuda et al [48]) are also used in VDC. The ABS speed sensor is used for obtaining vehicle speed (Bauer, [18]).

Signal conditioning of sensor signals is done on the sensor IC. As by-wire cars (Bretz [14]) slowly become reality, more sensors and electromechanical actuators will need to be used as parts of VDC systems. One example is steer by wire systems. The associated signal conditioning electronics is expected to be local to the actuator and the sensor. With the use of ever more sensors, actuators and ECUs soft and hard error diagnostics and fault tolerance in the event of a failure will become more important. According to Barron and Powers [30], a hard failure in an actuator such as a short circuit or an open circuit can easily be detected while soft sensor failures due to drifting voltage or current are not so easy to detect. For this reason, extensive double checking is done in VDC algorithms in an ECU to see if sensor readings are meaningful or not. Information from other sensors are used along with simple vehicle models for this type of error checking. In case of detection of such a fault, a fault tolerant controller that can work without that particular sensor is needed.

## 6. Man machine interface problems

Since the human driver is in the feedback loop in VDC, controller analysis, simulation and design should actually include a suitable human driver model. Note that one of the simplest models for a human driver is a high gain first order system with dead time which has a destabilizing effect on the overall VDC loop. The conventional uncontrolled car's lateral steering dynamics has the steering wheel rotary position as its input and the resulting yaw rate of the vehicle as its output. In the controlled car, the driver is within the feedback loop. The main idea of a VDC is to obtain the driver's intent by measuring the steering wheel command and to intervene if the vehicle does not follow the driver's command in an adequate fashion. To avoid stability problems due to man machine interaction, the VDC should use a fading effect, intervening only when necessary for durations of time shorter than the panic reaction time of the driver (Ackermann, [28]). This is called short duration intervention control or fading action control and has been used successfully by several researchers in the form of a fading integrator (Bünte [51], Ackermann [28]), in the form of a band pass filter (Aksun Güvenç et al [9]) and in the form of a limited integrator (Aksun Güvenç and Güvenç [6]).

## 7. Requirements on the ECU, networking requirements and implications for controller design

The electronic control unit (ECU) used for VDC algorithms has three times longer code than an ABS algorithm due to the need for extensive diagnostics/error checking algorithms. The preliminary testing of the VDC code in ECU level relies on HIL simulators for that purpose. In order to improve code development time and in order to reduce the risk of erroneous code in later stages of testing, the HIL simulator must be easily interfacable and must offer a highly realistic physical interface (in terms of electrical loads, network protocols and so on).

Currently, vehicles use several ECUs one for each demanding control task (one for engine control, one for ABS control, one for VDC etc) on a common network. This parallel mode of operation requires a high level of interoperability of these ECUs on the same network. The main IT related requirements on ECUs for VDC are faster, more reliable hardware and better network environments for control applications.

The Controller Area Network (CAN) of Bosch is a frequently used bus system in automotive control networks [15], [53]. The use of other network protocols is possible in the future (Lian et al, [16]). For control purposes, however, either a deterministic network like the CAN has to be used or if this approach is not taken, one has to deal with designing digital vehicle dyanmics controllers for nonuniform sampling. The use of a deterministic network is assumed here. In that case, the control designer must take into account sampled data operation of his controller at typically not so fast update rates. One further problem is that the yaw rate sensor signal will be available after a processing delay. To be more realistic, of course, one also needs to use a controller computational delay. To see the possible degradation effect due to the use of a networked approach, a linear single track model of a mid sized passenger car and a quite robust continuous time controller designed for it have been taken from the literature (Aksun Güvenç [7], Aksun Güvenç et al [9]). Simulation results have shown the degrading effect of different sampling rates and sensor reading delays on yaw moment disturbance rejection properties.

## 8. More recent work

More recent work that falls within the headings 2-7 above have become available after the initial preparation of this paper. The more recent work of the authors under these headings can be found in references [60]-[88].

## 9. Conclusions

Vehicle dynamics control is a very active area of research which is very promising for more capable algorithm development and for potential practical application. Each new capability places more demands on the information technology related software and hardware parts of VDCs. This paper identified several key subsections of a VDC with important IT related needs and constraints in a survey fashion.